\documentstyle[aps,prb,twocolumn,epsf]{revtex}

\begin{document}

\title{Broken-Symmetry States in Quantum Hall Superlattices}

\draft

\author{C. B. Hanna$^{1}$, J. C. D\'{\i}az-V\'elez$^{1}$,
    and A. H. MacDonald$^{2,3}$}
\address{$^{1}$Department of Physics, Boise State University,
Boise, Idaho, 83725}
\address{$^{2}$Department of Physics,
Indiana University, Bloomington, Indiana 47405}
\address{$^{3}$Department of Physics, University of Texas at Austin,
Austin, Texas, 78712}

\date{\today}

\maketitle

\begin{abstract}

We argue that broken-symmetry states with either spatially diagonal or
spatially off-diagonal order are likely in the quantum Hall regime, for
clean multiple quantum well (MQW) systems with small layer separations.
We find that for MQW systems, unlike bilayers, charge order
tends to be favored over spontaneous interlayer coherence.
We estimate the size of the interlayer tunneling amplitude needed to
stabilize superlattice Bloch minibands by comparing the variational
energies of interlayer-coherent superlattice miniband states with those
of states with charge order and states with no broken symmetries.
We predict that when coherent miniband ground states are stable,
strong interlayer electronic correlations will strongly enhance the
growth-direction tunneling conductance and promote the possibility
of Bloch oscillations.

\end{abstract}


\section{Introduction}

The strong-correlation physics of bilayer electron systems in
the quantum Hall regime has been of interest since shortly after
the discovery of the quantum Hall effect.\cite{halperin,dltheory}
Of particular interest is the occurrence of
broken-symmetry states with spontaneous interlayer phase
coherence\cite{reviews}(SILC) that continue to be a source of surprises
and puzzles.\cite{spielman,balents,stern,fogler,demler,schliemann}
In the quantum Hall regime, Landau-level degeneracy leads to competing
nascent broken symmetries; the SILC broken symmetry is driven by the
strong interlayer electronic correlations that it produces,
and competes subtly with a spatially diagonal broken symmetry state
in which electronic charge spontaneously occupies one of the two layers.
(In the case of a bilayer system with small interlayer separations and
low densities, the SILC state can occur in principle even at
zero magnetic field.\cite{zheng,hannaprb,sternsilc})

The energy of bilayer quantum Hall systems can be expanded
in powers of the layer separation $d$, the parameter
that most critically controls their properties.
The competition between SILC and charge-ordered states is decided
only by terms of second and higher order in $d$.
The property that SILC and charge-ordered states in bilayers have
the same energies to first order in $d$ is not an accident and can
be understood using the bilayer pseudospin language.\cite{dltheory}
In this description, electrons in top and bottom layers are eigenstates
of the $\hat z$ component of the pseudospin, while electrons with
interlayer coherence have pseudospin projections in the $xy$ plane.
Similarly, the charge-ordered and SILC-ordered states correspond,
respectively, to states with Ising and XY ferromagnetic order.
The competition between these two states depends on the 
sign of the pseudospin anisotropy energy.\cite{jungwirth}
The pseudospin-dependent part of the interaction in bilayers
is proportional to the difference between same-layer and different-layer
electron-electron interactions, which for two-dimensional electron-electron
interactions is proportional to $[1 -\exp(-qd)]/q$
in reciprocal space, where $q$ is the magnitude of the in-plane wave vector.
The origin of the mysterious absence of a pseudospin
anisotropy energy at first order in $d$ is now clear since the term
in this interaction that is first order in $d$ is independent of $q$;
{\it i.e.}, it is a delta-function interaction that has no effect on
fully spin-polarized fermions because of the Pauli exclusion principle.
The leading terms in the pseudospin magnetic anisotropy
energy\cite{jungwirth} in this case appears at second order in $d$ and,
as it turns out, leads to SILC rather than charge order.

In this paper, we generalize the investigation of order in quantum Hall
systems at integer total filling factors to the case of multilayers,
where the pseudospin analogy does not apply.  We should then expect the
energetic difference between charge (diagonal) and SILC (off-diagonal)
order to be settled at first order in $d$.  We find that, in contrast
to the bilayer case, charge order is favored in multilayers for small $d$.
States with interlayer phase coherence can, however, be stabilized by
relatively weak interlayer tunneling, and we predict that they will have
unusual physical properties.

Our conclusions are based on a comparison of variational energies for
states with charge order and states with SILC in
multilayer quantum Hall systems.\cite{multiaps}
The multilayer SILC states
are characterized by spontaneous finite-width Bloch minibands.
We compare the energies of these miniband wave functions
to those of states with strong but independent correlations
within each layer, including charge-ordered staggered states with
unequal layer densities.
Physically, this difference between bilayers and multiple-quantum-well
systems occurs because 
the Hartree energy per volume for states with a given charge imbalance
is far lower in the superlattice case, as much as four times lower
compared to bilayers.
In some cases however, relatively small interlayer tunneling amplitudes
are sufficient to stabilize uniform density miniband states.
We propose that these states could be identified experimentally by
a substantially enhanced growth-direction conductance that is due to their
strong interlayer correlations, and argue that they might also support
Bloch oscillations.

Multilayer quantum Hall systems have been fabricated which exhibit the
quantum Hall effect at integer filling factors per layer,\cite{multiint}
and an interesting body of recent work has focused on the 
chiral surface states that occur in this instance.\cite{chisurf,multiedge}
At fractional filling factors, most work\cite{joynt,kuramoto}
has concentrated on the physics at the special values of filling factor
per layer values for which Laughlin-Jastrow\cite{laughlin} states
with strong intralayer correlations occur.
Our interest here is in nearly disorder-free multilayer systems at strong
magnetic fields with general fractional filling factors per layer, $\nu$,
between zero and one.
Other recent work on multilayer quantum Hall systems at fractional
filling factor per layer has focused on the possible realization of
the Josephson effect,\cite{ezawa}
and on metal-insulator transitions\cite{multimi}
and critical exponents for localization in disordered systems.\cite{meir}
The possibility of charge order in superlattice quantum Hall systems
has been proposed previously,\cite{multith1,multith2,multith3} but
not confronted against competing SILC states.  Related physics
can in principle occur in other quasi two-dimensional electron systems,
for example layered organic conductors.\cite{mckenzie}

Our paper is organized as follows.
In Section \ref{sec:minigs} we present our SILC variational wave functions
and compare their energies with those of states which have neither
charge order nor interlayer correlations.
The SILC states, which have better interlayer correlations but less
favorable intralayer correlations, have lower energy when $d$ is small.
In Section \ref{sec:staggered} we consider charged-ordered
multiple-quantum-well states.
These states achieve the objective of reducing the liklihood of close
approaches between electrons in nearby layers simply by reducing the
average charge density in the neighboring layer, rather than by improving
interlayer correlations.  In this section we demonstrate that charge-ordered
states are favored over SILC states.
Section \ref{sec:transport} discusses the properties of the coherent
Bloch miniband states in the case where interlayer coherence is aided
by tunneling between the electron layers; although the coherence in this case
is not spontaneous, it is strongly enhanced by interactions.
We conclude in Section \ref{sec:conclusion}.

\section{Miniband Ground State}
\label{sec:minigs}

We consider a system with many coupled quantum wells separated by
a distance $d$, and a Landau-level filling factor per layer
$\nu \equiv N_e/(N_w N_{\phi})$ smaller than one.
Here $N_e,N_w,N_{\phi}$ are the total number of electrons,
the number of quantum wells containing two-dimensional electron layers,
and the number of flux quanta passing through each layer, respectively.
(The Landau-level degeneracy is $N_{\phi}$.)
Unless the interlayer separation is quite small
($d \sim \ell$, where $\ell$ is the magnetic length),
independent strongly correlated states will form in each two-dimensional
layer.  From theoretical and experimental studies of the fractional quantum
Hall effect, the energies of these states,\cite{fqheenergies} whose
character changes rapidly with $\nu$, is relatively accurately known.
In bilayer systems, as interactions between the two-dimensional layers 
increase in importance, the many-particle ground state develops
SILC, which promotes interlayer correlations at the cost of
partially disrupting correlations within the layers.
The variational wave function we propose to accomplish this compromise
in a superlattice is the following single Slater determinant:
\begin{equation}
\label{eq:varwf}
|\Psi_0\rangle = \prod_{q,X} c^{\dagger}_{q,X} |0\rangle.
\end{equation}
Here the single-particle miniband states are
\begin{equation}
|q,X\rangle = \frac{1}{\sqrt{N_w}} \sum_{j=1}^{N_w} e^{iqdj} |j,X\rangle ,
\end{equation}
where $|j,X\rangle$ is the state of a particle in the Landau-gauge
lowest-Landau-level state $|X\rangle$ in the $j$-th quantum well.
Such a miniband state has strong built-in interlayer correlations.
The variational wave function in Eq.~(\ref{eq:varwf}) is obtained from
the requirements that it have equal constant densities in each layer and 
have a density
matrix $\rho_{j_1j_2}(X)$ that is translationally invariant
(so that it depends only on the difference $|j_1-j_2|$ of layer indices):
\begin{equation}
\rho_{j_1j_2}(X) = \langle \Psi_0 | c_{j_1X}^\dagger c_{j_2X} | \Psi_0 \rangle
= \rho_0(|j_1-j_2|).
\end{equation}
This form for density matrix self-consistently solves
the  Hartree-Fock (HF) equations, and hence $|\Psi_0\rangle$
is a local extremum of the HF energy functional.

We focus here on the limit $N_w\rightarrow\infty$, appropriate to
a system with a large number of coupled quantum wells.
In Eq.~(\ref{eq:varwf}), the product $X$ goes over all $N_\phi$ states
within the lowest Landau level, but the product $q$ goes over only
a fraction $\nu$ of the growth-direction Bloch miniband states 
in the interval $-\pi/d<q\le\pi/d$.
As we discuss in Sec.~\ref{sec:transport},
the energy is minimized when the occupied wave vectors are contiguous,
but, in the absence of interlayer tunneling, is invariant
under a simultaneous translation of all occupied wave vectors.
For $N_w\rightarrow\infty$,
the ground-state density matrix for electrons separated by $j$ layers is
\begin{equation}
\rho_0(j) = \frac{\sin(\pi j \nu)}{\pi j} ,
\end{equation}
so that there are, in general, correlations between layers at all separations.

The miniband wave function in Eq.~(\ref{eq:varwf}) generalizes to
a multiple quantum well the notion of SILC studied previously in
bilayer\cite{dltheory,reviews} and
trilayer\cite{triplex,triplef,tripleh} systems.
This wave function will have a low energy at small interlayer separation
because it establishes correlations between electrons in different layers:
the pair distribution function for two electrons that are
$j$ layers apart and which have an in-plane separation $r$ is
\begin{equation}
g_j(r) = 1 - e^{-r^2/2\ell^2}
         \left[ \frac{\sin(\pi j \nu)}{(\pi j \nu)} \right]^2 .
\end{equation}
The decreased probability for finding pairs of electrons in
nearby layers with small in-plane separations 
lowers the energy when the layers are close together.

\begin{figure}[h]
\epsfxsize3.5in
\centerline{\epsffile{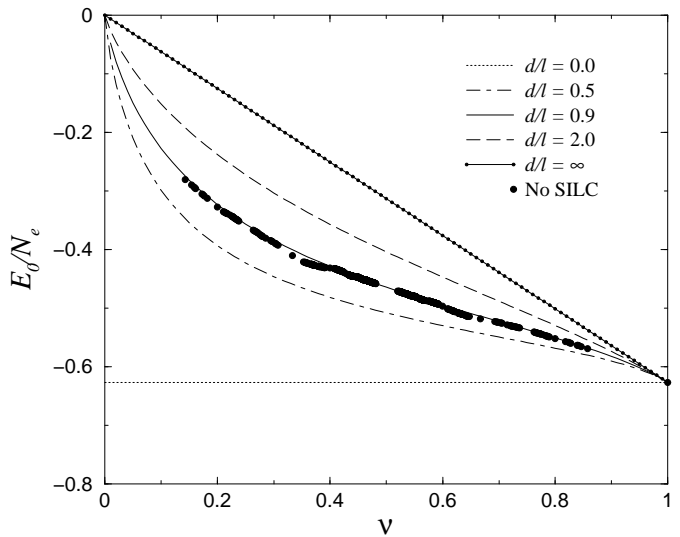}}
\caption{
Ground-state energies per particle, $\varepsilon(\nu)$,
in units of $e^2/4\pi\epsilon\ell$,
for an $N_w\rightarrow\infty$ miniband quantum Hall state
versus the filling factor per layer $\nu$, for $t$=0.
The estimated ground-state energies per particle for states
without correlations between the layers are shown as filled circles.
(``No SILC'').
}
\label{fig:enu}
\end{figure}

The HF ground-state energy per particle of the miniband state in
Eq.~(\ref{eq:varwf}) may be obtained by summing over the interactions
with correlation holes distributed over all layers:
\begin{eqnarray}
\label{eq:emini}
\varepsilon(\nu) &=& \frac{E_0(\nu)}{N_e} =
-2t \frac{\sin(\pi\nu)}{\pi\nu}
\\ \nonumber
&& - \frac{\nu}{2} \sum_{j=-\infty}^\infty
\int \! \frac{d^2r}{2\pi\ell^2} \;
\left[ 1 - g_j(r) \right]
\frac{e^2/4\pi\epsilon}{\sqrt{r^2 + (dj)^2}} ,
\end{eqnarray}
where $-t$ is the interlayer tunneling matrix element in the
tight-binding approximation, $e$ is the electronic charge,
and $\epsilon$ is the dielectric constant of the host semiconductor.
Note that interlayer tunneling ($t>0$) lowers the energy of the
miniband state, and that when this term is present, the interlayer coherence
in $|\Psi_0\rangle$ is not spontaneously generated.  In Fig.~\ref{fig:enu},
we plot the ground-state energy per particle, $\varepsilon$=$E_0/N_e$,
versus $\nu$ for the miniband quantum Hall state at $t$=0
for different layer separations, to estimate the energy per particle
of a uniform-density superlattice ground state possessing SILC.
We also plot the estimated energy per particle for isolated-layer states
without SILC.\cite{isolayer}
For $d < d_c \sim 0.9\ell $, the superlattice state with
SILC has a lower energy than independent-layer states with identical
layer densities.
The calculations used to produce Fig.~\ref{fig:enu} would seem to indicate
that $\nu$=1/2 is the most favorable filling factor per layer for
observing SILC.
This is because the relatively high energy per particle of the $\nu$=1/2
independent-layer state leads to a larger critical distance $d_c$ 
below which the $t$=0 miniband state is stable.
Conversely, $\nu$=1/3 would appear to be among the least-favored
filling factors for obtaining SILC, because the sizeable negative 
correlation energy per particle of the $\nu$=1/3 independent-layer
Laughlin states\cite{kuramoto} gives a smaller value of $d_c$.

The comparison of the energies of variational wave functions depicted
in Fig.~\ref{fig:enu} shows that the $t$=0 (SILC) miniband state
has a lower energy than identical uniform layers of independent states
when $d<d_c\sim 0.9\ell$.
In the case of balanced bilayer systems with filling factor
$\nu$=1/2 per layer, the energy of the Slater-determinant variational
SILC state is lower than that of two $\nu$=1/2 states without
interlayer correlations for $d/\ell < d_c/\ell \sim 1.16$.
This result for $d_c$ for balanced bilayers is in close agreement with
other theoretical estimates:
the time-dependent HF collective-mode stability criterion\cite{ahm}
gives $d_c/\ell$=1.18, and finite-size exact diagonalization
studies\cite{schliemann} give $d_c/\ell$=1.175.
This is consistent with experimental findings\cite{murphy,spielman}
when the finite-thickness\cite{schliemann} of the electrons layers
and quantum fluctuations\cite{yogesh} are taken into account.
In double-quantum-well systems, theoretical predictions\cite{fertig,ahm}
of SILC have been confirmed experimentally, most directly in recent
experiments which have discovered a Josephson-like peak in the interlayer
tunneling conductance.\cite{spielman}
Thus Fig.~\ref{fig:enu}, taken at face value, is evidence for the possibility
of a miniband SILC state at sufficiently small layer separation.
However, as explained in the introduction, the competition between SILC
and charge order is a subtle one even in the bilayer case, and there is good
reason to expect that this competition plays out differently in a
multiple-quantum-well system.
We therefore turn our attention in the following section to variational
wave functions with charge order.

\section{Staggered States}
\label{sec:staggered}

We showed in the previous section that for sufficiently small interlayer
separations, the miniband state is always energetically favored over
independent-layer states of equal uniform density.
However, we must also consider the possibility that at small interlayer
separations, the exchange interaction could favor independent-layer states
with unequal densities.
Indeed, the  possibility of staging transitions in multilayer systems was
discussed in Ref.~\onlinecite{multith2}, where HF calculations predicted
a series of staging transitions in which occupied layers were separated by
an increasing number of empty layers as the particle density and interlayer
separation were decreased.
In the bilayer case, the predicted\cite{rudenwu} staging transition
is not expected to occur, even at zero magnetic
field;\cite{zheng,hannaprb,sternsilc}
instead, it is preempted by a transition to a bilayer SILC state
with equal layer densities.\cite{reviews}
Even in the case of the bilayer quantum Hall state, the SILC state is
barely favored over the unblanced state with $\nu$=1 in one layer
and $\nu$=0 in the other layer for $d/\ell \rightarrow 0$:
the two states are degenerate at $d$=0 and, as explained in the 
introduction, the SILC state is lower
in energy only at order $d^2$.  The bilayer SILC state is most stable
at intermediate densities between 0 and $d_c$.

In the superlattice case, we show below that the ($t$=0) miniband SILC
state loses energetically to staggered states at small $d$.
One important reason for this is that the Hartree energy per volume of a
staggered-density state is far lower in a superlattice than in a bilayer.
This is understood qualitatively by considering the case of
two-dimensional charge sheets of alternating areal charge density $\pm\sigma$.
For a superlattice, the bulk-system requirement that the voltage drop
due to the electric fields be zero across two layer spacings
produces an electric field of constant magnitude that alternates in sign
across each layer: $E_\infty=\pm \sigma/2\epsilon$, from Gauss' Law.
The same calculation for a classical bilayer capacitor gives an
electric field that is twice as large, $E_2=\sigma/\epsilon$.
The Hartree energy per unit volume is given by $\epsilon E^2/2$,
so it costs four times less energy per volume to have a staggered state
in a superlattice as compared to a bilayer.

To make the comparison more precise, we have computed the energy 
of the SILC state analytically in the limit $d\rightarrow 0$
by summing up the interactions of an electron with exchange hole
contributions from remote layers.  The two states compete by measuring the 
loss in exchange energy in the SILC state that is due to 
removing a part of the exchange hole to remote layers
against the Hartree energy of the charge-ordered state discussed above.  
{}From Eq.~(\ref{eq:emini}), we find that for the SILC state,
\begin{equation}
\label{eq:esmalld}
\varepsilon(\nu) \rightarrow
\varepsilon(1) \left[ 1
- \frac{d/\ell}{4\nu} \left(\frac{2}{\pi}\right)^{5/2} \ln(\ell/d)
+ \frac{d(1-\nu)}{2\ell} \right]
\end{equation}
for $d \rightarrow 0$, where
$\varepsilon(1)$ = $\varepsilon(\nu$=$1)$ =
$-(1/2) \sqrt{\pi/2} \: e^2/4\pi\epsilon\ell$.
Because of the $d\ln(\ell/d)$ term above, the SILC state always has
a higher energy in the limit $d\rightarrow 0$ than a staggered state
that alternates filled ($\nu$=1) layers with completely empty layers,
since the Hartree cost of such staggered states grows only linearly with $d$.
However, the HF approximation underestimates the miniband energy,
and it may be that correlation effects (and finite-thickness effects)
improve the energetics of the miniband states.
Unfortunately, we cannot provide a reliable estimate of the
correlation-energy contribution to the miniband state,
since the generalized random-phase approximation
gives a logarithmically divergent (and negative) result.\cite{yogesh}
We are not able to make a definitive conclusion on the possibility that
fluctuation effects
beyond the HF approximation stabilize the miniband state
for small or even zero interlayer tunneling, or whether there are
other superlattice SILC states besides the miniband state which could
be realized, leaving this as an issue that must ultimately be decided 
experimentally.  In Section \ref{sec:transport} we discuss
experimental signatures of the miniband state.

It is useful to compare the energy of the miniband state
to those of more general staggered states.
Consider a staging transition of order $n$ to a staggered state
that consists of repeating units of $n$+1 layers,
with one layer having filling factor $\nu$+$n\delta$
and $n$ depleted layers having filling factor $\nu$-$\delta$.
The driving force for the staging transition is
the interlayer exchange-correlation interaction,
which favors concentrating the electrons in a single plane.
The main energy cost of the staging transtions is the Hartree charging
energy due to having charge-imbalanced layers.
We note that the surface charge density of the
$n$ depleted layers is $e\delta/2\pi\ell^2$, so that the Hartree contribution
to the energy per particle for the staggered states is\cite{multith2}
\begin{equation}
\varepsilon_H(n) = \frac{n(n+2)}{12} \frac{e^2}{4\pi\epsilon\ell}
\frac{\delta^2}{\nu} \frac{d}{\ell} ,
\end{equation}
which is always linear in $d$.
At large separations, the Hartree cost
for the staging transitions is prohibitive, so $n$=0.
The exchange-correlation contribution to the total energy per particle
for independent-layer states is
\begin{equation}
\varepsilon_{xc}(n) = \frac{
(\nu + n\delta) \varepsilon_1(\nu + n\delta)
+ n(\nu - \delta) \varepsilon_1(\nu - \delta)}{(n+1)\nu} ,
\end{equation}
where $\varepsilon_1(\nu)$ is the energy per particle for a single-layer
quantum Hall state at filling factor $\nu$, which is indicated by the
dots in Fig.~\ref{fig:enu}.  Note that $\varepsilon_1(\nu)$ decreases
with increasing $\nu$, which favors increasing the density of some layers
at the expense of others.
The total energy per particle of the staggered system is
$\varepsilon_s$=$\varepsilon_H$+$\varepsilon_{xc}$.
For the sake of definiteness,
we consider two special cases below, $\nu$=1/2 and $\nu$=1/4.

\begin{figure}[h]
\epsfxsize3.5in
\centerline{\epsffile{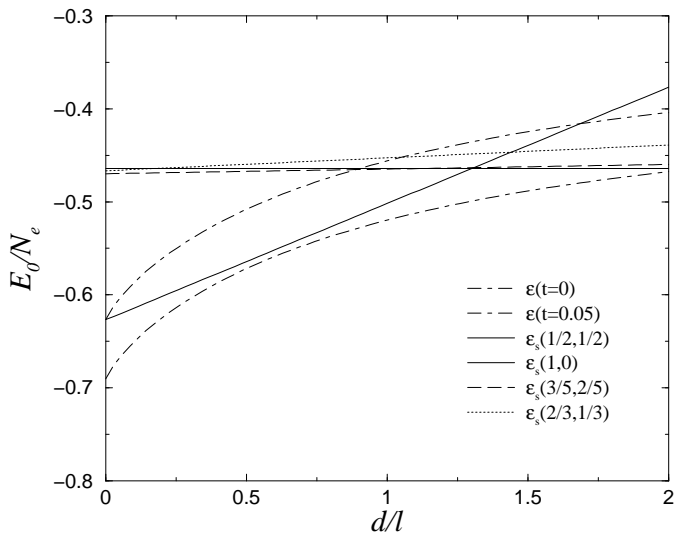}}
\caption{
Ground-state energies per particle, $\varepsilon$, in units of
$e^2/4\pi\epsilon\ell$,
for an $N_w\rightarrow\infty$ miniband quantum Hall state
at $\nu$=1/2 versus the layer separation $d/\ell$.
The estimated ground-state energies per particle $\varepsilon_s$
for staggered states without correlations between the layers are also shown.
Interlayer tunneling of order $t \sim 0.05 \: e^2/4\pi\epsilon\ell$
is required to stabilize the miniband state.
}
\label{fig:ehalf}
\end{figure}

For $\nu$=1/2, we consider an $n=1$ independent-layer staggered state with
$\delta$=$(1/2)/(2k-1)$ for $k$=1,2,3,$\infty$, corresponding to a
superlattice state with a two-layer unit cell with filling factors (1,0),
(2/3,1/3), (2/5,3/5) and (1/2,1/2), respectively.
The energy per particle for these states is plotted in Fig.~\ref{fig:ehalf},
along with the energy per particle for the miniband state at $t$=0 and
$t$=0.05 (in units of $e^2/4\pi\epsilon\ell$.)

Figure \ref{fig:ehalf} shows the energy per particle at $\nu$=1/2 for
miniband states (dashed-dot curves) at $t$=0 (upper) and $t$=0.05 (lower),
for $n$=1 staggered independent-layer states consisting of pairs of
layers with alternating filling factors (1,0) (sloped solid line),
(2/3,1/3) (dotted line) and (3/5,2/5) (dashed line), and for the
$n$=0 independent-layer state with $\nu$=1/2 (horizontal solid line).
Note that for $t$=0, the $n$=1 staggered state (1,0) has the lowest
energy for $d/\ell < 1.3$, after which the $n$=0 independent
layer state with $\nu$=1/2 has the lowest energy.  Thus the miniband
state requires finite interlayer tunneling to be stabilized: for
$t$=0.05, the miniband energy is always lower than that of the staggered (1,0)
state, and is lower than that of the $n$=0 independent-layer state for
$d/\ell < 2$.  Note that finite interlayer tunneling produces a constant
downward shift of the miniband energy at fixed $\nu$ and $t$, but is
expected to have a much smaller effect (of order $t^2$, from perturbation
theory) on the independent-layer states.

\begin{figure}[h]
\epsfxsize3.5in
\centerline{\epsffile{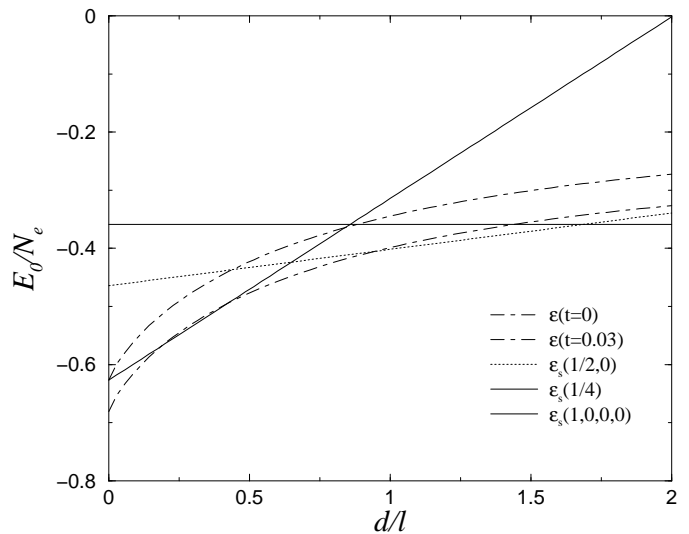}}
\caption{
Ground-state energies per particle, $\varepsilon$,
in units of $e^2/4\pi\epsilon\ell$,
for an $N_w\rightarrow\infty$ miniband quantum Hall state
at $\nu$=1/4 versus the layer separation $d/\ell$.
The estimated ground-state energies per particle $\varepsilon_s$
for staggered states without correlations between the layers are also shown.
Interlayer tunneling of order $t \sim 0.03 \: e^2/4\pi\epsilon\ell$
is required to stabilize the miniband state at small $d/\ell$.
}
\label{fig:equarter}
\end{figure}

For $\nu$=1/4, we consider an $n$=3 independent-layer staggered state
corresponding to a superlattice state with a four-layer unit cell
with filling factors (1,0,0,0), and a two-layer unit cell
with filling factors (1/2,0).  The energy per particle $\varepsilon_s$
for these states is plotted in Fig.~\ref{fig:equarter}, along with the
energy per particle, $\varepsilon$, for the miniband state at
$t$=0 and $t$=0.03 (in units of $e^2/4\pi\epsilon\ell$).

Figure \ref{fig:equarter} shows the energy per particle at $\nu$=1/4 for
miniband states (dashed-dot curves) at $t$=0 (upper) and $t$=0.03 (lower),
for an $n$=3 staggered independent-layer states consisting of four
layers with alternating filling factors (1,0,0,0) (sloped solid line),
for an $n$=1 staggered independent-layer states consisting of pairs of
layers with alternating filling factors (1/2,0) (dotted line),
and for the $n$=0 independent-layer state with $\nu$=1/4 (horizontal
solid line).
Note that for $t$=0, the $n$=3 staggered state (1,0,0,0) has the lowest
energy for $d/\ell < 0.6$, after which the $n$=1 staggered state (1/2,0)
has a lower energy until $d/\ell \approx 1.6$, after which the $n$=0
independent-layer state with $\nu$=1/4 has the lowest energy.
At $\nu$=1/4, the miniband state requires finite interlayer tunneling
of order $t \sim 0.03$ to be stabilized.
When the miniband state is stabilized by tunneling, the
interlayer correlations of the miniband state are expected to have
a strong effect on growth-direction transport, as we discuss in
the next section.

\section{Interlayer Transport}
\label{sec:transport}

Individual particle transport between correlated electron layers in
independent-layer quantum Hall states is strongly suppressed\cite{oldtunnel}
because many-particle states containing
the electron and hole created by a tunneling event 
have small overlaps with low-energy states.
Miniband states can have a large conductivity
because of the possibility of collective transport, much as the 
tunneling conductance in bilayer systems is increased by many orders
of magnitude\cite{spielman} when interlayer coherence is established. 

The nature of disorder in multiple-quantum-well systems
is important for transport considerations, and indeed for the
formation of the miniband state.  Unlike double-quantum-well
systems which can be modulation-doped from the sides,
multiple-quantum-well systems must have layers of dopants between the
quantum wells.  These layers will create disorder within the
quantum wells and in the interlayer tunneling amplitude.
In the bilayer case, it is expected theoretically,
and known experimentally, that interlayer correlations occur
only in weakly disordered systems.  The occurrence of the physics we
propose will require, in all likelihood, special efforts to limit
disorder due to  modulation doping between the layers. One possible
strategy is to place the interlayer dopants in a deep quantum well,
creating carriers that can screen lateral disorder but which do not
contribute to growth-direction transport.\cite{lp}
The following discussion assumes that conditions with
relatively modest disorder can be achieved.

To describe coherent transport, we separate the tunneling amplitude
between the layers into a nonrandom part which is diagonal in
Landau-level state indices, $-t$, and a fluctuating part,
$\delta t_{X',X}$. In the absence of disorder and interactions,
a constant tunneling amplitude would lead to a Bloch miniband with dispersion
$\varepsilon_0(q)$=$-2t\cos(qd)$. In a semiclassical approximation,
the effect of an electric field ${\cal E}$ on the spontaneous miniband
state is to move it rigidly in the reduced-zone $k$-space at the
rate $\dot K$=$-e {\cal E}/\hbar$.  When $\delta t$ and the disorder within
the layers are ignored, the macroscopic current density carried by
the miniband state is given by
\begin{equation}
j = \frac{e}{\hbar} \frac{\nu}{2\pi\ell^2 d}
\frac{\partial \varepsilon_K}{\partial K}
\end{equation}
where $\varepsilon_K$ is the  energy per particle for the miniband state
when the center of the occupied region is located at $K$ in $k$-space.
It is easy to show that only the band energy $\varepsilon_0(q)$
contributes to the $K$ dependence of $E_{\rm tot}$ so that
\begin{equation}
\label{eq:jbloch}
j = \frac{e}{\hbar} \frac{\nu}{2\pi\ell^2} \frac{\sin(\pi\nu)}{\pi\nu}
2t \sin(eVT/\hbar) ,
\end{equation}
where $V$=${\cal E}d$ is the interlayer voltage difference,
and $T$ in Eq.~(\ref{eq:jbloch}) denotes time.
These time-dependent oscillating currents are just the Bloch
oscillations\cite{blochorig} that are expected to occur for
noninteracting electrons in a low-disorder limit that has never
been approached in any degenerate electron system. Related
effects have, however, been seen in multiple quantum wells
with optically excited carriers.\cite{blochopt}
The role of strong interactions and miniband formation
in the quantum Hall regime is to allow this physics to appear in
samples where the transport at zero magnetic field is incoherent.

\begin{figure}[h]
\epsfxsize3.5in
\centerline{\epsffile{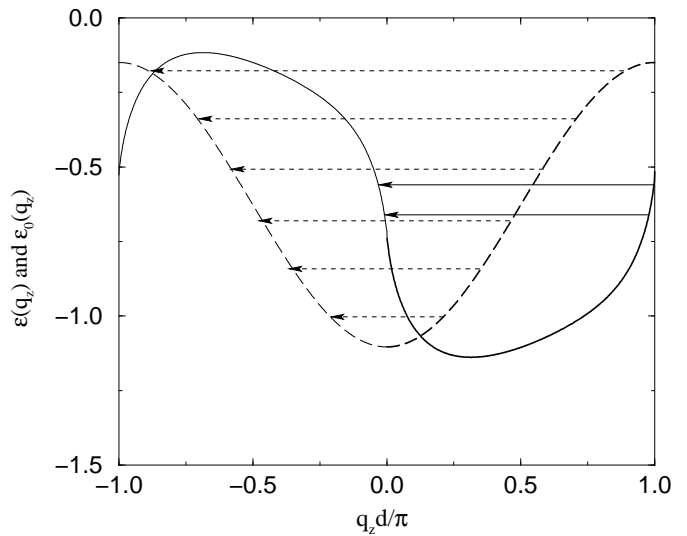}}
\caption{
The bare single-particle dispersion $\varepsilon_0(q)$=$-2t\cos(qd)$
for $t$=$0.24 \: e^2/4\pi\epsilon\ell$ (dashed),
and the HF dispersion $\varepsilon(q)$ for $\nu$=1/2 and
$t$=$0.05 \: e^2/4\pi\epsilon\ell$ (solid).
The dark portions of the curves, on the right-hand side, indicate
the occupied states, whose momenta we have shifted to the right
by $K$=$\pi/2d$ to produce a current-carrying state.
The miniband states shift to the right with $K$,
and have a much smaller phase space for scattering.
}
\label{fig:eshift}
\end{figure}

That interactions support the robustness of the disorder-free
quasiparticle bands can be seen in Fig.~\ref{fig:eshift}, where we plot
the quasiparticle bands for
the maximum-current state with $K$=$\pi/2d$,
for $t$=$0.05 \: e^2/4\pi\epsilon\ell$ and $\nu$=1/2.
The HF eigenvalues are given by
\begin{equation}
\label{eq:ehf}
\varepsilon(q) = -2t\cos(qd)
- \frac{1}{N_w} \sum_{p} f(q-p) X(p) ,
\end{equation}
where $f(q)$=$\langle c^\dagger_{qX} c_{qX} \rangle$ is the
wave-vector occupation function, and $X(p)$ is the HF self-energy,
\begin{equation}
X(p) = \frac{e^2}{4\pi\epsilon\ell} \int_0^\infty dx e^{-x^2/2}
\left[ \frac{\sinh(xd/\ell)}{\cosh(xd/\ell)-\cos(pd)} \right] .
\end{equation}
Equation~(\ref{eq:ehf}) shows that if there were no interlayer tunneling,
the quasiparticle bands would shift rigidly with $K$.
For non-interacting electrons, shifting the occupied momenta by
$K$=$\pi/2d$ would open up an enormous phase space for scattering
(dashed arrows of Fig.~\ref{fig:eshift}),
because for every occupied state, there would be an unoccupied state
of opposite momentum.
Compared to the interaction-free case, very few of the occupied
states in the nonequilibrium finite-$K$ state are degenerate with
the unoccupied quasiparticle states (solid arrows of Fig.~\ref{fig:eshift}).
Scattering is further reduced because the available scattering states
are all near $k_F$,
where the density of states is greatly suppressed,
even for the maximum current-carrying state.
The system is {\em not} a superconductor
because the quasiparticles in the current-carrying
state are {\em not} in equilibrium; nevertheless, the phase space
for current relaxation by disorder scattering is immensely reduced.

The basic physics that we believe will lead to suppressed current
relaxation in miniband quantum Hall states is explained in the
above paragraph.
The linear conductivity in the growth direction can be evaluated
more formally, following the lines of a recent calculation of bilayer 
interlayer conductance.\cite{joglekar}  When applied to superlattices,
these considerations imply that the conductance is 
approximately that of the auxiliary single-particle system with 
interaction-enhanced growth-direction quasiparticle velocities and 
suppressed quasiparticle scattering rates.  This suggests that the
growth-direction conductivity will be enhanced by orders of 
magnitude upon entering the miniband state.

\section{Conclusion}
\label{sec:conclusion}

We have investigated the stability of uniform-density interlayer-coherent
quantum Hall miniband states, comparing their energies to those of states
without interlayer correlations, including staggered states with diagonal 
order.  Our calculations demonstrate that SILC does not occur in superlattices
within the HF approximation, although 
coherent miniband states with strongly enhanced band widths can be stabilized 
by weak interlayer tunneling.   For small interlayer separations,
independent-layer staggered states with diagonal order are preferred
because they have larger exchange energies and because the  
Hartree energy cost of inhomogeneity is not large: the Hartree energy per
volume for a staggered state can be four times lower in a superlattice
as compared to a bilayer.
The superlattice case differs in this regard from that of a bilayer,
which does exhibit SILC.  At large layer separations, intralayer
correlations become more important energetically than interlayer correlations,
and independent-layer states of uniform density and constant filling factor
are favored, just as for bilayers.

The quantum Hall miniband state, when formed (e.g., with the help of
interlayer tunneling), is expected to have a strongly enhanced
growth-direction conductivity.  This is because, in the miniband state as
in the SILC bilayer state, electrons in different layers are arranged
so as to accomodate tunneling electrons from other layers.
This greatly increased growth-direction conductivity may be the most
definitive experimental signature of the superlattice miniband state.
We emphasize that the tunneling conductance we describe is strongly enhanced
by interlayer exchange and correlations over the value expected by
single-particle tunneling alone.  

We also suggested that a quantum Hall miniband state would constitute a
promising candidate for a Bloch oscillator, without the need for optically
excited carriers.  The period of Bloch oscillations is
$T_B$=$2\pi\hbar/(e{\cal E}d)$ for an applied DC electric field
${\cal E}$,\cite{blochorig}
where $d$ is the distance between the layers.  For typical samples
at room temperature, the scattering relaxation time is about $10^9$ times
smaller than the Bloch period $T_B$, which precludes the possibility
of observing Bloch oscillations.
Quantum Hall miniband states are expected to have greatly reduced
scattering rates because of their many-body interlayer correlations.

We caution that our proposals must be regarded as uncertain, since they are 
based on single-Slater-determinant variational wave functions.  For example,
interlayer correlations can be established by quantum fluctuations not
included in this HF theory without breaking any symmetries.
Additionally, the fact that interlayer correlations establish a charge gap
in bilayer systems, but not in multiple-quantum-well systems, may make 
mean-field-theory considerations less reliable in the present case.
We therefore feel that the true nature of the ground state can only
be established experimentally and, with this in mind, have argued that 
enhanced growth-direction conductivity due to collective transport 
is a reliable signature of spontaneous or interaction-enhanced interlayer
coherence.

Aside from a large miniband-enhanced growth-direction conductivity
and the possibility of producing Bloch oscillations,
there may also be other many-body effects that are analogous
to those proposed for bilayer SILC systems.\cite{reviews}
These includes novel effects produced by in-plane magnetic fields,
enhanced Coulomb drag at zero temperature,
and the existence of charged topological excitations.
The collective modes of miniband states are very different
in character from those of a superlattice system with
independent-layer states.\cite{multith1,multith2,multith3}
The collective-mode spectrum of the miniband state could be explored
by tunneling conductivity measurements made in a parallel magnetic
field,\cite{tunnelmode} or by resonant inelastic light
scattering.\cite{pinczuk}

We also note that while finite tunneling is needed to stabilize the
miniband state within the HF approximation, it is still an open
question whether SILC becomes favorable when quantum fluctuations,
which lower the miniband energy, are included.  Finite thickness is also
known to enhance SILC in bilayer systems.\cite{schliemann}
It is also likely that the uniform (in-plane) density staggered states
that we have considered have a higher energy than states with nonuniform
densities in the plane.
For example, the (1,0) staggered state may break up into domains
of $\nu$=1 and $\nu$=0 within a plane, so that the average filling factor
of the plane remains $\nu$=1/2.  This would reduce the Hartree energy
without too large a sacrifice in exchange energy.  Such a nonuniform state
might also possess interlayer correlations at the boundaries
between $\nu$=1 and $\nu$=0 regions.  
It is also possible that at some filling factors, in particular those at
which Laughlin-Jastrow states occur, other types of states
with SILC could be realized, including states that possess both charge
order and SILC.  While this work is necessarily incomplete
in exploring all of these possibilities, and unable to reach definitive 
conclusions concerning the possibilities that have been explored, it does
demonstrate that the interaction physics of multiple-quantum-well systems
at fractional filling factors per layer
will be even richer than that of single-layer systems,
if systems with sufficiently weak disorder can be fabricated.

\acknowledgements

This research was supported by the Welch Foundation,
by the National Science Foundation under grants DMR-9972332
and DMR-0115947,
and by a grant from the Research Corporation.
CBH is thankful for the support provided by the ITP Scholars Program
at the Institute for Theoretical Physics at U.C. Santa Barbara,
where part of this work was carried out.
This research was supported in part by the National Science Foundation
under Grant No. PHY99-07949.
The authors gratefully acknowledge the hospitality of
St. Francis Xavier University and helpful interactions with L. F. Stones.

\end{document}